\documentclass[12pt,preprint]{aastex}

\shorttitle{Coupled calculation of tidal heating of extrasolar planet}
\shortauthors{Shoji et al.}

\begin{document}

\title{Thermal--orbital coupled tidal heating and habitability of Martian-sized extrasolar planets around M stars} 


\author{D. Shoji and K. Kurita}
\affil{Earthquake Research Institute, University of Tokyo}




\begin{abstract}
M type stars are good targets in the search for habitable extrasolar planets. Because of their low effective temperatures, the habitable zone of M stars is very close to the star itself. For planets close to their stars, tidal heating plays an important role in thermal and orbital evolutions, especially when the planet orbit has a relatively large eccentricity. Although tidal heating interacts with the thermal state and orbit of the planet, such coupled calculations for extrasolar planets around M star have not been conducted. We perform coupled calculations using simple structural and orbital models, and analyze the thermal state and habitability of a terrestrial planet. Considering this planet to be Martian sized, the tide heats up and partially melts the mantle, maintaining an equilibrium state if the mass of the star is less than 0.2 times the mass of the Sun and the initial eccentricity of the orbit is more than 0.2. The reduction of heat dissipation due to the melted mantle allows the planet to stay in the habitable zone for more than 10 Gyr even though the orbital distance is small. The surface heat flux at the equilibrium state is between that of Mars and Io. The thermal state of the planet mainly depends on the initial value of the eccentricity and the mass of the star. 
\end{abstract}



\section{Introduction}
The search for extrasolar planets has revealed that there are many terrestrial planets out of our solar system (e.g., L\'{e}ger et al. 2009). By detecting these extrasolar terrestrial planets, many researches have tried to determine the planets' thermal and orbital histories (Henning et al. 2009; Barnes et al. 2010; L\'{e}ger et al. 2011; Wagner et al. 2012). The thermal and orbital states of extrasolar terrestrial planets are important for the planets' evolution. In addition, the thermal and orbital states affect the habitability, making analysis of these factors of key importance. 

Recently, in the search for habitable extrasolar planets, M stars were suggested for observation (Scalo et al. 2007) because the number of M stars is much larger than that of G stars. Conventionally, M stars are not considered to be likely hosts of habitable planets because their planets' orbits are too close, which results in the planets being tidally rocked (synchronously rotated) and their atmosphere freezing on the dark side. However, Joshi et al. (1997) hypothesized that carbon dioxide can make a planet habitable. Thus, the importance of M star extrasolar planet analysis continues to increase.

Focusing on the terrestrial planets in our solar system, the main heat sources are the insolation from the Sun and the radioactive decay in the planets themselves. However, if extrasolar planets orbit around small stars such as an M star, the tidal heating has a significant effect on the habitability. Compared with the Sun, M stars have low temperatures, and thus the magnitude of the insolation becomes low. Therefore, the habitable zone is closer to the star. The rate of change of a planet's orbit due to tidal dissipation increases with decreasing orbital distance. For this reason, a planet orbiting an M star may not be able to stay within the habitable zone for a long time. In addition, assuming that the planets orbit according to Kepler's law, the tidal frequency is high, producing tidal heating that may exceed the radiogenic and insolation heating. In contrast to the terrestrial planets in our solar system, such tidal heating is an important factor for the thermal and orbital evolutions of planets orbiting M stars.

The effects of tidal heating on terrestrial planets orbiting M stars have been analyzed (Jackson et al. 2008; Barnes et al. 2008). Although these researches revealed that tidal heating influences a planet habitability, they assumed that the magnitude of the dissipation of the planet is constant. The tidal heating mechanism is complicated and interacts with the internal thermal state and orbit of the planet (e.g., Hussmann \& Spohn 2004). When a planet is tidally heated, structure and thermal states of the planet changes, which affects the tidal heating by the feed-back. In addition, the orbit and heating rates also interact with each other. Tidal heat dissipation is dependent on the eccentricity of the orbit; the eccentricity decreases with  heating rate because orbital energy is lost as tidal heat. Thus, tidal heating should be coupled with orbital and thermal evolution. 

In this work, we evaluate the coupled calculation of tidal heating considering the thermal and orbital states. However, coupled calculations are very complex. In addition, the internal structure of extrasolar terrestrial planets is not constrained well.  Thus, as a fist step, we consider a Martian-sized terrestrial planet. An advantage of this choice is that the internal pressure of planets of this size is small compared with earth-sized planets, due to their small radii, which allows a simple model of the internal structure to be used. A Martian-sized planet has been discovered (Muirhead et al. 2012); however, this planet is outside the habitable zone. Despite this, we can expect similarly sized planets to be discovered within habitable zones in future observations. Considering the observation, Earth-sized planets are easier to detect compared to Martian-sized planet. However, we try the coupling theoretical calculation with Martian structure. In our solar system, Io is a tidally locked rocky satellite with high tidal heating (Segatz et al. 1988). Thus, we reference the thermal model of Io as well as Mars. If accurate internal structures of super-earths are derived in the future, we can apply the results of this work to such models with more complicated structures.

\section{Theory}
\subsection{Structure}
\begin{figure}[htbp]
\centering
\includegraphics[width=13cm,angle=-90] {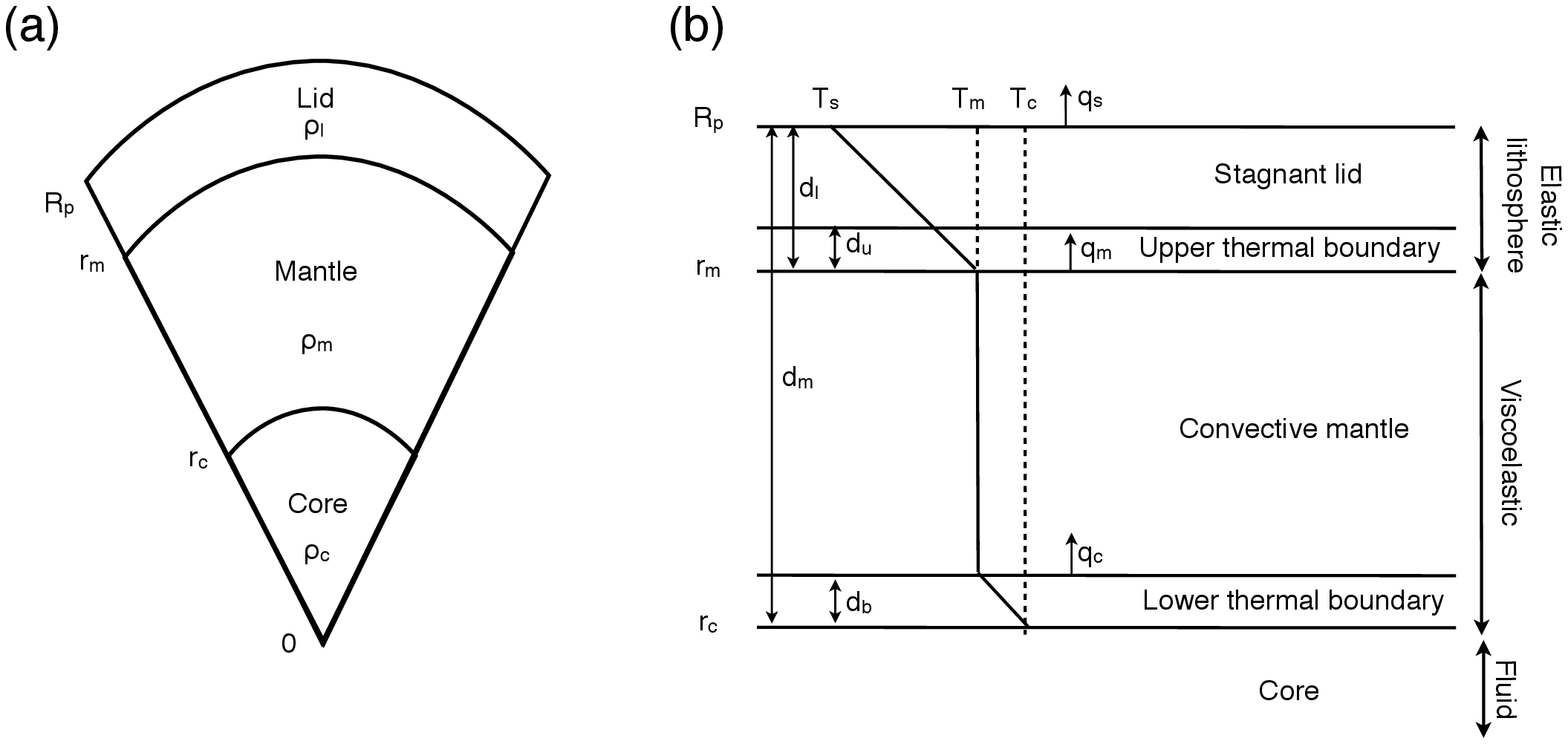}
\caption{Structural and thermal models of the planet at (a): global scale and (b): local scale. For the calculations, the stagnant lid and upper thermal boundary of the mantle are assumed to be elastic because of the low temperature (defined as the lithosphere). Tidal heating occurs in the convective mantle and lower thermal boundary area. The core is fluid, and no tidal heating occurs in the core.}
\label{structure}
\end{figure}

In this work, for simplicity, we assume that the internal structure of the planet is composed of three layers: a liquid core, a mantle and a stagnant lid (Fig. \ref{structure} a). Stagnant lid convection occurs in the mantle. Because stagnant lid convection is the most typical convection system, scaling laws which relate the magnitude of the convection to the structure have been analyzed by many researches (e.g., Reese et al. 1999 b). Although mobile lid convection (e.g., O'Neill \& Lenardic 2007) or heat pipe state (Moore 2001; Moore \& Webb 2013) are suggested as alternative heat transfer mechanisms, we assume the stagnant lid convection as the heat transfer model in this work. Rheology and temperature distributions depend only on the radial distance (i.e., are independent of latitude and longitude). The mantle is differentiated into three parts: two thermal boundaries and a well mixed convective section (Fig. \ref{structure} b). At the two boundaries, heat is transferred by conduction. In this work, we define the lid and the upper boundary of the mantle as the lithosphere. The lithosphere is assumed to be elastic, and tidal heat is generated only in the mantle by viscoelastic dissipation (Hussmann \& Spohn 2004). This assumption is valid because the surface temperature is less than 300 K in the habitable zone, which is much smaller than the mantle temperature (more than 1600 K). In the convective part, for simplicity, we assume that convective temperature is $T_m$ and independent of depth. $T_m$ changes with time both by the heating and cooling.

The rate of change of the mantle and core temperatures $T_m$ and $T_c$ are calculated based on the energy balance (Breuer \& Spohn 2006).

\begin{equation}
\rho_m c_m V_m\epsilon_m(St+1)\frac{dT_m}{dt}=-4\pi r_m^2q_m+4\pi r_c^2q_c+Q_{\mathrm{tide}}+Q_{\mathrm{rad}}
\label{Tm}
\end{equation}

\begin{equation}
\rho_c c_c V_c \epsilon_c\frac{dT_c}{dt}=-4\pi r_c^2q_c.
\label{Tc}
\end{equation}
Here, the subscript $m$ and $c$ refer to properties of the mantle and core, respectively. $\rho_m$ and $\rho_c$ are the densities,  $c_m$ and $c_c$ are the heat capacities, $V_m$ and $V_c$ are the volumes, $r_m$ and $r_c$ are the radial distances from the center to the convective mantle and the core, respectively. $\epsilon_{m}$ is the ratio between the average mantle and upper boundary temperatures (Grott \& Breuer 2008 b). {\bf In the case of adiabatic thermal model, temperature changes with the depth. Thus, effect of $\epsilon_m$ is required. $\epsilon_m$ ranges between 1.0 and 1.05 (Breuer \& Spohn 2006; Grott \& Breuer 2008 b). Because this range is around 1.0, Eq. (\ref{Tm}) is applicable to the isothermal mantle model which we consider in this work. We conducted the calculations with different values of $\epsilon_m$, and confirmed that the value of $\epsilon_{m}$ little affects the results.} $\epsilon_{c}$ is the ratio between the average core and lower boundary temperatures, which is 1.1 (Breuer \& Spohn 2006; Grott \& Breuer 2008 a). The values of each parameter are shown in Table \ref{value}. $St$ is the Stefan number, which accounts for the consumption or release of latent heat by the mantle (Breuer \& Spohn 2006). $St$ is represented as the ratio of latent heat to the heat capacity of the mantle, and is given by

\begin{equation}
St=\frac{L_m}{c_m}\frac{V_a}{V_m}\frac{d\phi}{dT_m},
\label{stefan}
\end{equation}
where $L_m$ is the latent heat of the mantle (Breuer \& Spohn 2006). $\phi$ is the {\bf volume} melt fraction, which is discussed below. $V_a$ is the volume of the melt zone. {\bf Effect of melt in the rock has been studied by many researches (e.g., Rosenberg \& Handy 2005). When mantle begins to melt, melt is distributed to the frame and junctions of grain boundaries. By increasing the melt fraction, melts among grains begin to be interconnected each other. However, as long as the melt fraction is under the critical value called the rheological critical melt percentage (RCMP), grain framework is maintained (Rosenberg \& Handy 2005). When the melt fraction exceeds the RCMP, melt covers the grain completely and effect of liquid dominates. Considering these effects, if the melt fraction is under the RCMP, we assume that the convective area of the mantle is well mixed and that melt is not separated from the solid mantle; thus, $V_a=V_m$.} This assumption may not be valid if the melt fraction exceeds the RCMP because the generated melt segregates from the solid mantle. However, we found that melt fraction of our coupled calculation is less than the 25 \% at every condition, and Moore (2003) estimates that the segregation of melt occurs at the 40 \% to 60 \% of melt fraction. Thus the assumption of mantle with partial melt state is reasonable. $Q_{\mathrm{rad}}$ is the heating due to radioactive decay. For simplicity, we use the averaged radiogenic heating in the Martian mantle estimated by Grott \& Breuer (2008 b). The volumetric radiogenic heating $q_{\mathrm{rad}}$ is given by 

\begin{equation}
q_{\mathrm{rad}}=q_0\exp(-\lambda t),
\label{q_rad}
\end{equation}
where $q_0=1.6\times10^{-8}$ W m$^{-3}$ and $\lambda=1.5\times10^{-17}$ s$^{-1}$ (Grott \& Breuer 2008 b). Thus, $Q_{\mathrm{rad}}=q_{\mathrm{rad}}V_m$. We do not consider the radiogenic heating from other parts of the planet.
In the core of the Earth, potassium can be the large heat sources in the core, and Mars may contain potassium, but the stability of potassium in iron is extremely small at low pressure such as the case for the Martian core (Gessmann \& Wood 2002). Thus, we ignored the heat generation in the core.

{\bf $q_m$ is the heat flux from the mantle, which is given by (Nimmo \& Stevenson 2000). In the case of stagnant lid convection, $q_m$ can be related to the Nusselt number $Nu$ as}

\begin{equation}
q_m=k_m\frac{\Delta T}{d_m}Nu,
\label{flux}
\end{equation}
{\bf where $k_m$ is the thermal conductivity of the mantle. $\Delta T$ is the temperature difference between top and bottom of the mantle. $d_m$ is the thickness of the mantle. Assuming the newtonian viscosity, $Nu$ can be scaled by the Rayleigh number $Ra$ and Frank-Kamenetskii parameter $\theta$ (Reese et al. 1999 a) as}

\begin{equation}
Nu\approx0.5\theta^{-\frac{4}{3}}Ra^{\frac{1}{3}}.
\label{nusselt}
\end{equation}
{\bf $Ra$ is given by}

\begin{equation}
Ra=\frac{\rho_m g \alpha \Delta T}{\kappa \eta_m}d_m^3,
\label{rayleigh}
\end{equation}
{\bf where $g$, $\alpha$ and $\kappa$ are acceleration due to gravity, thermal expansivity and thermal diffusivity of the mantle, respectively. $\eta_m$ is the viscosity of the mantle, which is discussed in the next section. In this work, we assume the Arrhenius type viscosity, which mainly depends on the temperature. $\theta$ is given by}

\begin{equation}
\theta=\Delta T\frac{E}{RT_i^2},
\label{theta}
\end{equation}
{\bf where $E$ and $R$ are the activation energy and the gas constant, respectively. $T_i$ is the internal temperature and defined as the mean temperature at the upper thermal boundary (Reese et al. 1999 b). Using Eqs. (\ref{flux})-(\ref{theta}) $q_m$ is given (Nimmo \& Stevenson 2000) as}

\begin{equation}
q_m=\frac{k_m}{2}\left(\frac{\rho_m g \alpha}{\kappa \eta_m}\right)^{\frac{1}{3}} \gamma^{-\frac{4}{3}},
\label{qm}
\end{equation}
where  

\begin{equation}
\gamma=\frac{E}{RT_i^2}.
\label{gamma}
\end{equation}
Using the typical value of $T_i$=1800 K and $E=300$ kJ/mol, $\gamma$ becomes 0.011 (Nimmo \& Stevenson 2000). In order to calculate the accurate value of $\gamma$, numerical calculation of convection is required (Reese et al. 1999 b). However, in this work, we fixed $\gamma$ at 0.011 for simplicity. The heat flux from the core to the mantle $q_c$ is given by

\begin{equation}
q_c=k_m\frac{T_c-T_m}{d_b},
\label{heat_c}
\end{equation}
 where $d_b$ is the thickness of the lower boundary. 
 
 Thickness of the upper and lower thermal boundaries ($d_u$ and $d_b$) are derived by Nimmo \& Stevenson (2000). Through the upper thermal boundary, temperature drops by 8/$\gamma$ and $q_m$ of heat flux is transported from the mantle to the surface (Nimmo \& Stevenson 2000). They mention that 8/$\gamma$ of temperature drop is found by Solomatov (1995). Thus, $d_u$ is given by 
 
 \begin{equation}
d_u=\frac{8k_m}{\gamma q_m}.
\label{d_u}
\end{equation}
 In the case of the thickness of the lower boundary $d_b$, they assume that local Rayleigh number of the lower thermal boundary is the same to that of the upper thermal boundary. Under this assumption, local Rayleigh numbers are related as
 
 \begin{equation}
 \frac{\rho_m g\alpha(8/\gamma)d_u^3}{e^4\eta_m\kappa}=\frac{\rho_m g\alpha(T_c-T_m)d_b^3}{\exp(-0.5\gamma[T_c-T_m])\kappa\eta_m}=Ra_c,
 \label{du_db}
 \end{equation}
 where $e$ and $Ra_c$ are the Napier's constant and critical Rayleigh number (Nimmo \& Stevenson 2000). From Eq. (\ref{du_db}), the ratio between $d_b$ and $d_u$ is given (Nimmo \& Stevenson 2000) as
 
 \begin{equation}
 \frac{d_b}{d_u}=0.5(\gamma[T_c-T_m])^{-1/3}\exp\left(-\frac{\gamma(T_c-T_m)}{6}\right).
 \label{ratio}
 \end{equation}
 $d_b$ can be calculated from Eqs. (\ref{d_u}) and (\ref{ratio}).
 The thickness of the lithosphere $d_l$ and the surface heat flux $q_s$ are related by

\begin{equation}
q_s=k_{l}\frac{T_m-T_s}{d_l},
\label{qs}
\end{equation}
where $k_{l}$ and $T_s$ are the thermal conductivity of the lithosphere and the surface temperature, respectively. In this work, we assume that tidal and radiogenic heatings are generated in the mantle. {\bf Heat generation and loss in the lithosphere are ignored.} Under this assumption, $q_s$ equals $q_m$. Thus, from Eqs (\ref{qm}) and (\ref{qs}), the lithosphere thickness $d_l$ can be calculated, which is required for the calculation of tidal heating.

If the surface of the planet is assumed to be in equilibrium with the insolation heat flux uniformly distributed over the surface, the surface temperature $T_s$ is given by (L\'{e}ger et al. 2009; Barnes et al. 2010)

\begin{equation}
T_s=(1-A)^{1/4}g_e\left(\frac{R_{\ast}}{2a}\right)^{1/2}T_{\ast}.
\label{Ts}
\end{equation}
If the heat transport is limited to the light side,

\begin{equation}
T_s=(1-A)^{1/4}g_e\left(\frac{R_{\ast}}{a}\right)^{1/2}T_{\ast}.
\label{Ts_one}
\end{equation}
$A$ is the albedo of the planet, which is assumed to be constant at 0.3 (Paige et al. 1994). $g_e=1.0$ is the effectiveness of the heat transport. $a$ is the orbital distance, which is discussed in Sec. 2.3. $R_{\ast}$ is the radius of the star, which can be related to the mass of the star through

\begin{equation}
\log_{10}\frac{R_{\ast}}{R_{\odot}}=1.03\log_{10}\frac{M_{\ast}}{M_{\odot}}+0.1,
\label{R_st}
\end{equation}
where $M_{\odot}$ and $R_{\odot}$ are the mass and radius of the Sun, respectively (Gorda \& Svechnikov 1999). $M_{\ast}$ is the mass of the star. $T_{\ast}$ in Eq. (\ref{Ts}) is the effective temperature of the star, which is given by 

\begin{equation}
T_{\ast}=\left(\frac{L}{4\pi\sigma R_{\ast}^2}\right)^{1/4},
\label{T_st}
\end{equation}
where $\sigma$ and $L$ are the Stefan-Boltzman constant and luminosity of the star, respectively (Barnes et al. 2008). $L$ is related to the mass of the star by 

\begin{equation}
\log_{10}\frac{L}{L_{\odot}}=4.101\mu^3+8.162\mu^2+7.108\mu+0.065
\label{L_st},
\end{equation}
where $\mu=\log_{10}(M_{\ast}/M_{\odot})$ (Scalo et al. 2007). $L_{\odot}$ is the luminosity of the Sun. Using Eqs. (\ref{Ts})-(\ref{L_st}), $T_s$ can be determined. We use the constant values as shown in Table \ref{value} except for $M_{\ast}$ and $a$. $a$ changes with time according to the evolution of the orbit. In the case of $M_{\ast}$, we conduct the calculations by varying $M_{\ast}/M_{\odot}$ from 0.1 to 0.5, which is a reasonable range for an M star. 

Melt fraction $\phi$ depends on the solidus and liquidus temperatures. The melt fraction is assumed to be 1.0 at the liquidus temperature and increases linearly from the solidus to liquidus temperature (Moore 2003). The solidus $T_{\mathrm{sol}}$ and liquidus $T_{\mathrm{liq}}$ temperatures of the mantle are given as functions of pressure $P$ (Takahashi 1990):  

\begin{equation}
T_{\mathrm{sol}}=1409.15+134.2P-6.581P^2+0.1054P^3~~\mathrm{K}
\label{T_sol}
\end{equation}

\begin{equation}
T_{\mathrm{liq}}=2035+57.46P-3.487P^2+0.0769P^3 ~~\mathrm{K}.
\label{T_liq}
\end{equation}
Pressure depends on the depth $z$ and can be calculated as 

\begin{equation}
P(z)=\int^{z}_{R_p}\rho(z)g(z)dz,
\label{P}
\end{equation}
where $R_p$ is the radius of the planet. $\rho(z)$ is the density at $z$. $g(z)$ is the acceleration due to gravity at $z$, which can be expressed as

\begin{equation}
g(z)=\frac{M(r)G}{r^2},
\label{gravity}
\end{equation}
where $r$ and $G$ are radial distance ($r$=$R_p$--$z$) and the gravitational constant, respectively. $M(r)$ is the mass enclosed by a sphere of radius $r$, which is

\begin{equation}
M(r)=\int^r_04\pi r'^2\rho(r') dr'.
\label{M_r}
\end{equation}
We assume that the solidus and liquidus temperatures of the mantle are averaged between the top and bottom of the mantle (Hussmann \& Spohn 2004):

\begin{equation}
\bar{T}_{\mathrm{sol}}=\frac{T_{\mathrm{sol}}(r_m)+T_{\mathrm{sol}}(r_c)}{2}
\label{T_sol_ave}
\end{equation}

\begin{equation}
\bar{T}_{\mathrm{liq}}=\frac{T_{\mathrm{liq}}(r_m)+T_{\mathrm{liq}}(r_c)}{2}.
\label{T_liq_ave}
\end{equation}
When $T_m$ exceeds $\bar{T}_{\mathrm{sol}}$, the mantle begins to melt. Thus, $\phi$ can be expressed as

\begin{equation}
\phi=\frac{T_m-\bar{T}_{\mathrm{sol}}}{\bar{T}_{\mathrm{liq}}-\bar{T}_{\mathrm{sol}}}.
\label{melt_fraction}
\end{equation}

\subsection{Tidal heating}
For the calculation of the tidal heating, the mantle is assumed to behave viscoelastically. The tidal heating rate $Q_{\mathrm{tide}}$ can be obtained as follows (Segatz et al. 1988).

\begin{equation}
Q_{\mathrm{tide}}=-\frac{21}{2}\frac{(R_p n)^5e^2}{G}\mathrm{Im}(\tilde{k}_2)
\label{tide}
\end{equation}
$n$ is the mean motion of the planet. Assuming that the planet orbits according to Kepler's law, $n$ can be related to the orbital distance $a$ by $n=(GM_{\ast}/a^3)^{1/2}$. $e$ is the eccentricity of the orbit, which is important for the coupled calculation because it interacts with the tidal heating. The evolution of $e$ is discussed in Sec. 2.3. $\tilde{k}_2$ is the second degree complex Love number, which largely depends on the internal structure and rheology of the planet. In the case of spherical-shell structured models, the complex Love number can be calculated using the methods based on the gravity potential of the planet (Segatz et al. 1988; Tobie et al. 2005; Roberts \& Nimmo 2008). Detailed theory and the precise method are reviewed by Tobie et al. (2005) and Roberts \& Nimmo (2008). Here, we mention the outline.

Assuming the spherical-shell structure, displacement, stress and potential can be separated into radial function part and spherical harmonics (Alterman et al. 1959; Takeuchi \& Saito 1972). Assuming the radial functions of displacement ($y_1$ and $y_2$), stress ($y_3$ and $y_4$), and gravity potential ($y_5$ and $y_6$) of a spherical shell model, they are related each other via propagation matrix $A_{ij}$ as 

\begin{equation}
\frac{dy_i}{dr}=A_{ij}y_j
\end{equation}
where $r$ is the radial distance from the center of the planet. For bodies with liquid cores, Sabadini \& Vermeersen (2004) derived the components of $A_{ij}$, which we use in this work. Each radial function is solved by the boundary conditions at the core and the surface. Love number of the spherical shell structure can be calculated (Tobie et al. 2005) by 

\begin{equation}
k_2=-y_5(R_p)-1.
\label{Love}
\end{equation}
Note that Love number in Eq. (\ref{Love}) is not complex number because the radial function theory is developed for the elastic body. However, viscoelastic rheology model can be related to the elastic model by Fourier transformation, which is called correspondence principle (Bilot, 1954). By the Fourier transformation, radial functions of viscoelastic model becomes complex value in the domain of the frequency of applied force. Complex radial functions also can be related linearly via propagation matrix although each component becomes complex number. By solving the radial functions by the viscoelastic model, complex Love number can be calculated as

\begin{equation}
\tilde{k}_2=-\tilde{y}_5(R_p)-1,
\label{Love_complex}
\end{equation}
where $\tilde{y}_5$ is the complex radial function of potential.

As for the rheology, we consider the Andrade model. An Andrade body is proven to fit well to the experimental results for the response of mantle rock (Jackson et al. 2002). The creep function for an Andrade body $J(t)$ is given by 

\begin{equation}
J(t)=\frac{1}{\mu_m}+\beta t^{n_a}+\frac{t}{\eta_m},
\label{creep}
\end{equation}
where $t$, $\mu_m$ and $\eta_m$ are time, the shear modulus and viscosity of the mantle, respectively (Jackson et al. 2002; McCarthy \& Castillo-Rogez 2013). $n_a$ and $\beta$ are the experimental parameters, which represent the effects of the anelasticity of the material such as heterogeneity of the grain. If $\beta$ is zero, the creep function becomes the conventional viscoelastic rheology (Maxwell model). $\beta$ can be expressed as (Castillo-Rogez et al. 2011) 
 
\begin{equation}
\beta=\frac{\mu_m^{n_a-1}}{\eta_m^{n_a}}.
\label{beta}
\end{equation}
This relationship is valid if the dissipation is dominated mainly by diffusion creep (Castillo-Rogez et al. 2011). $n_a$ is the unconstrained empirical parameter, which ranges from 0.2 to 0.5. In this work, we use $n_a=0.25$, which is based on the laboratory expedient of olivine (Jackson et al. 2002). Creep function at the frequency domain can be given (McCarthy \& Castillo-Rogez 2013) by the Fourier transformation as 

\begin{equation}
\tilde{J}(\omega)=\frac{1}{\mu_m}+\omega^{-n_a}\beta\cos\left(\frac{n_a\pi}{2}\right)\Gamma(n_a+1)-i\left[\frac{1}{\eta_m\omega}+\omega^{-n_a}\beta\sin\left(\frac{n_a\pi}{2}\right)\Gamma(n_a+1)\right],
\label{creep_complex}
\end{equation}
where $\omega$ is the frequency, which is the same to the mean motion in the case of tidal heating. $\Gamma$ and $i$ are the gamma function and $\sqrt{-1}$, respectively. Complex shear modulus $\tilde{\mu}(\omega)$ of the Andrade model can be calculated as 1/$\tilde{J}(\omega)$. Propagation matrix $A_{ij}$ contains the complex shear modulus, which affects the tidal dissipation of the planet.

In the calculation of the tidal heating, the shear modulus and viscosity of the mantle are the most important factors. At temperatures under $\bar{T}_{\mathrm{sol}}$, the shear modulus is assumed to be constant at 50 GPa, which is a typical value for mantle rock. As for the viscosity, we adopt an Arrhenius function, which is the typical description for the mantle viscosity (e.g., Breuer \& Spohn 2006): 

\begin{equation}
\eta_m=\eta_0\exp\left(\frac{E}{R}\left[\frac{1}{T_m}-\frac{1}{T_0}\right]\right).
\label{eta}
\end{equation}
$T_0=1600$ K is the reference temperature (Breuer \& Spohn 2006) and $\eta_0$ is the reference viscosity at $T_0$.  

When the temperature becomes higher than $\bar{T}_{\mathrm{sol}}$, the mantle begins to melt. 
If the melt is contained within the mantle, the shear modulus and viscosity drop drastically. As mentioned above, we assume that the partially molten mantle is well mixed. Reduction of the shear modulus and viscosity is controlled by the melt fraction $\phi$. Moore (2003) gives the viscosity over the solidus temperature. When the melt fraction $\phi$ is under RCMP (40\% to 60\%), the viscosity of the mantle is multiplied by $\exp(-B\phi)$ where $B$ is the dimensionless melt fraction coefficient. 
Thus, $\eta_m$ becomes

\begin{equation}
\eta_m=\eta_0\exp\left(\frac{E}{R}\left[\frac{1}{T_m}-\frac{1}{T_0}\right]\right)\exp(-B\phi).
\label{eta_melt}
\end{equation}
 In this work, we assume that RCMP is 40 $\%$. If the melt fraction exceeds the disaggregation point, the viscosity follows the Rascoe-Einstein relationship, which can be written (Moore 2003) as 
 
\begin{equation}
\eta_m=10^{-7}\exp\left(\frac{4\times10^4 ~\mathrm{K}}{T_m}\right)(1.35\phi-0.35)^{-5/2}  ~~\mathrm{Pa~ s}.
\label{eta_aggre}
\end{equation}
The shear modulus over the solidus temperature was determined by Fischer and Spohn (1990) to be 
$\mu_m=10^{(\mu_1/T_m+\mu_2)}$ Pa,
where $\mu_1=8.2\times10^4$ K. They also found that $\mu_2=-40.6$ due to the continuous condition at the solidus temperature (1600 K in their work). In our calculation, the solidus temperature was calculated using Eq. (\ref{T_sol_ave}). Thus, the shear modulus over the solidus temperature is 

\begin{equation}
\mu_m=10^{[\mu_1/(\Delta T_m+1600 \mathrm{K})+\mu_2]} ~~\mathrm{Pa},
\label{mu_melt}
\end{equation}
where $\Delta T_m=T_m-\bar{T}_{\mathrm{sol}}$. In addition, if $\phi$ exceeds the disaggregation point (0.4), the shear modulus is assumed to drop to $10^{-7}$ Pa (Moore 2003). 

\begin{figure}[htbp]
\centering
\includegraphics[width=13cm] {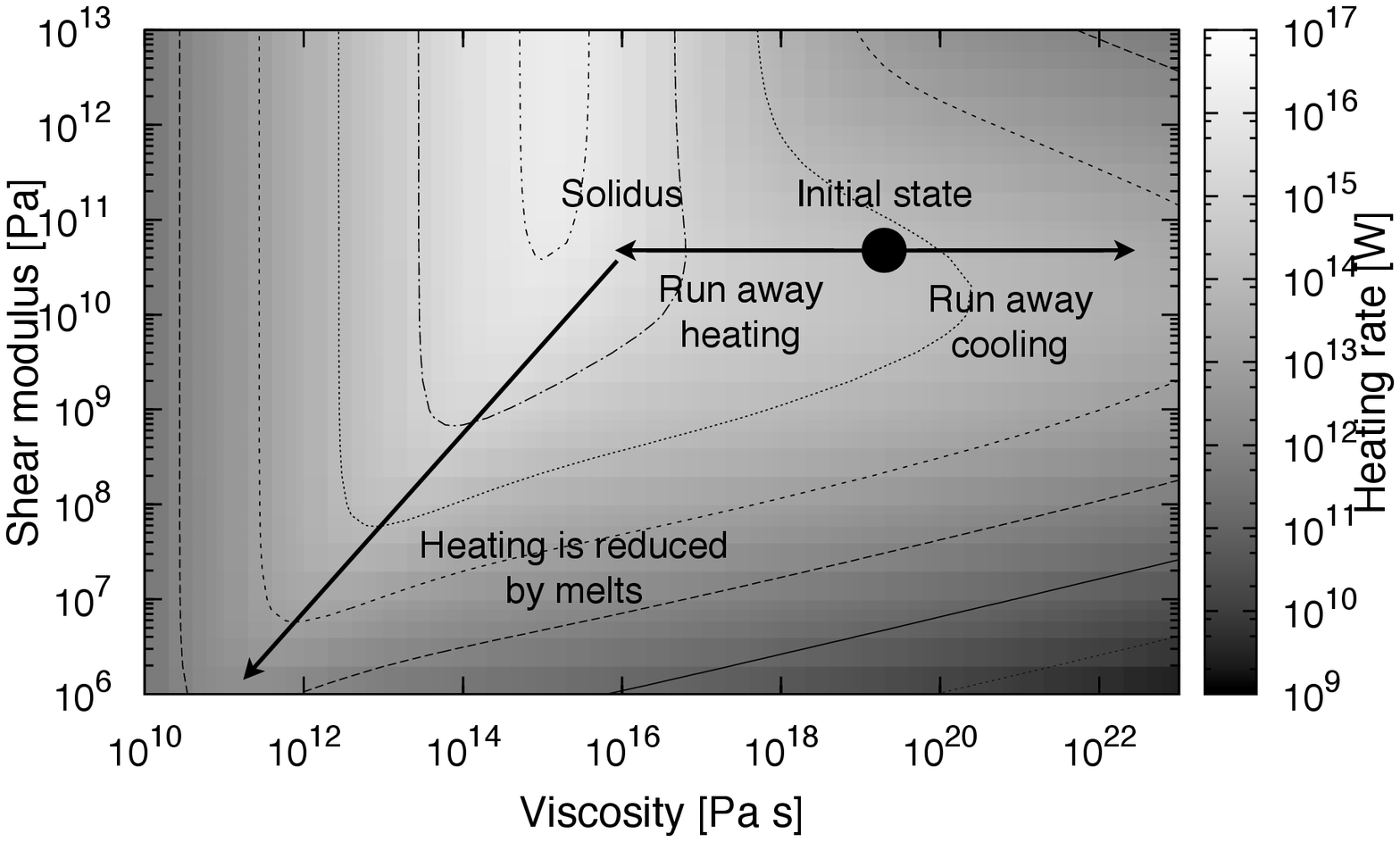}
\caption{Heating rate of the two layer (core+mantle) model as a function of shear modulus and viscosity of the mantle. $M_{\ast}$ and $e$ are $0.1M_{\odot}$ and 0.2, respectively. Assuming the initial state at the circle point, two types of heating evolution (runaway heating or cooling) can be expected. If the temperature exceeds the solidus temperature, the melt reduces the mantle shear modulus and viscosity, which results in decreased tidal heating and ends runaway heating.}
\label{heat}
\end{figure}

Figure \ref{heat} shows the tidal heating rate in the case of two layers (liquid core and mantle) as a function of viscosity and shear modulus of the mantle. Heating rates are calculated by Eqs. (\ref{tide})-(\ref{creep_complex}) with $M_{\ast}=0.1M_{\odot}$ and $e=0.2$. In the case of a 50 GPa shear modulus, the heating rates are maximized at a viscosity of $10^{15}$ Pa s. If we assume that the initial viscosity is $10^{19}$ Pa s, two types of heating state are possible. One evolution type is runaway cooling. If the heat radiated from the surface is larger than the tidal heating, the mantle cools and the viscosity increases. When the viscosity increases, the heating rates decrease as shown in Fig. \ref{heat}. Thus, mantle cools more.  The other heating state is runaway heating. If the tidal heating is large, the mantle temperature increases and the viscosity decreases according to Eq. (\ref{eta}), which results in increased tidal dissipation. However, runaway heating stops when the solidus temperature is reached. Because of the melt, the shear modulus and viscosity decrease, which results in a reduction of the heating rate (Fig. \ref{heat}). 

The coupled calculations in this work consider the three-layer model (the elastic lithosphere is included). The lithosphere works to decrease the heating rate because the displacement is reduced. An elastic layer can be included by considering the large viscosity of a viscoelastic layer. In our calculations, we set the viscosity of the lithosphere to $10^{40}$ Pa s. It is important to set a characteristic time for the viscoelastic body ($\eta/\mu$) that is sufficiently larger than the tidal period to produce the desired elastic behavior.  The shear modulus is set to 50 GPa. Thus, a viscosity of $10^{40}$ Pa s is sufficient to approximate the elastic body. 

\subsection{Orbital change }
The interaction between tidal heating and orbital evolution is very complicated. In this work, following the work by Barnes et al. (2008), we use the classical orbital equations, which are based on the works of Goldreich \& Soter (1966). The rates of change of orbital distance $a$ and eccentricity $e$ are as follows.

\begin{equation}
\frac{da}{dt}=-\left(21\frac{\sqrt{GM_{\ast}^3}R_p^5k_2}{M_pQ_p}e^2+\frac{9}{2}\frac{\sqrt{G/M_{\ast}}R_{\ast}^5M_p}{Q'_{\ast}}\right)a^{-11/2}
\label{a}
\end{equation}

\begin{equation}
\frac{de}{dt}=-\left(\frac{21}{2}\frac{\sqrt{GM_{\ast}^3}R_p^5k_2}{M_pQ_p}+\frac{171}{16}\frac{\sqrt{G/M_{\ast}}R_{\ast}^5M_p}{Q'_{\ast}}\right)a^{-13/2}e
\label{e}
\end{equation}
Here $Q_p$ and $Q'_{\ast}$ are the Q values of the planet and the star. Note that $Q'_{\ast}$ is divided by two thirds and includes the value of the Love number of the star (Barnes et al. 2008). The first terms represent the effect of the  dissipation of the planet. The second terms show the dissipation of the star. $k_2$ is the Love number of the planet, which is not complex but a real number. In the case of the viscoelastic rheology model, $k_2$ can be approximated by $|\tilde{k}_2|$. In addition $Q_p$ can be represented by Re($\tilde{k}_2$)/Im($\tilde{k}_2$). The imaginary part of Love number is much smaller than the real part. Thus, we can write $k_2/Q_p\approx\mathrm{Im}(\tilde{k}_2)$. The complex Love number depends on the internal structure and thermal state of the planet, which affects the heating rate as shown in Eq. (\ref{tide}). Using the approximation shown above, we can relate the orbital, thermal, and heating evolution of the planet. We have to consider that Eqs. (\ref{a}) and (\ref{e}) are not valid when the orbital period of the planet is shorter than the rotational period of the star. In this case, the signs of the second terms change (Barnes et al. 2008). However, due to the large mass of the star compared to the Martian-sized planet, the values of the second terms are significantly smaller than those of the first terms. Thus, Eqs. (\ref{a}) and (\ref{e}) are valid in this work.  If the planet's mass is relatively large we need to consider the rotational and orbital periods, which may be important for the case of a super-earth. $Q'_{\ast}$ represents the magnitude of the star's dissipation, which also affects the orbital evolution of the planet. We tried the calculation with values between $10^5$ and $10^7$. However, even at $Q'_{\ast}=10^{5}$, the first terms are larger than the second terms by more than six orders of magnitude. Thus, $Q'_{\ast}$ does not affect the orbital evolution of the planet, and the effect of the dissipation of the planet is dominant unless $Q'_{\ast}$ is unreasonably small. Thus, we use the constant value of $Q'_{\ast}=10^{5.5}$ in every calculation shown in this paper.

The purpose of this work is a habitability analysis through coupled tidal calculations. Thus, as an initial condition for the orbital distance, we set the center of the planet between the inner and outer edges of the habitable zone (Jackson et al. 2008). The inner and outer edges of the habitable zone ($l_{in}$ and $l_{out}$, respectively) were calculated by Barnes et al. (2008):

\begin{equation}
l_{in}=(l_{in\odot}-a_{in}T'_{\ast}-b_{in}T'^2_{\ast})\left(\frac{L}{L_{\odot}}\right)^{1/2}(1-e^2)^{-1/4}
\label{l_in}
\end{equation}

\begin{equation}
l_{out}=(l_{out\odot}-a_{out}T'_{\ast}-b_{out}T'^2_{\ast})\left(\frac{L}{L_{\odot}}\right)^{1/2}(1-e^2)^{-1/4},
\label{l_out}
\end{equation}
where $a_{in}=2.7619\times10^{-5}$ AU K$^{-1}$, $b_{in}=3.8095\times10^{-9}$ AU K$^{-2}$, $a_{out}=1.3786\times10^{-4}$ AU K$^{-1}$ and $b_{out}=1.4286\times10^{-9}$ AU K$^{-2}$. $T'_{\ast}=T_{\ast}-5700$ K. $T_{\ast}$ can be calculated using Eq. (\ref{T_st}). However, if $T_{\ast}$ is less than 3700 K, it should be 3700 K when considering $T'_{\ast}$ (Barnes et al. 2008).  $l_{in\odot}$ and $l_{out\odot}$ are the inner and outer edges of the solar habitable zone, respectively. They depend on the condition of cloud.  $l_{in\odot}$ is $\sim$0.89 AU, $\sim$0.72 AU and $\sim$0.49 AU when the cloud cover is 0\%, 50\% and 100\%, respectively. $l_{out\odot}$ is $\sim$1.67 AU, $\sim$1.95 AU and $\sim$2.4 AU when the cloud cover is 0\%, 50\% and 100\%, respectively. 

In this work, we take into account the habitable zone estimated by Barnes et al. (2008) as shown above. However, habitable zone shown in Eqs. (\ref{l_in}) and (\ref{l_out}) is based on the conventional estimations by Kasting et al. (1993) and Selsis et al. (2007). Recently, Kopparapu et al. (2013) revised the habitable zone based on new datas. In addition, habitable zone changes with time due to the evolution of the central star. We discuss the effect of the estimation and evolution of the habitable zone in the next section.

\subsection{Calculation procedure}
Although many equations are required for the coupled calculation, Eqs. (\ref{Tm}), (\ref{Tc}), (\ref{tide}), (\ref{a}) and (\ref{e}) are the main equations needed. Once the initial values of $T_m$, $T_c$, $e$ and $a$ are set, we can calculate properties of the internal structure such as the thickness of each layer and the thermal properties such as the surface temperature of the planet. Using the determined conditions, the complex Love number can be calculated, which is needed to calculate the tidal heating rate (Eq. (\ref{tide})) and orbital evolution (Eqs. (\ref{a}) and (\ref{e})). Using the calculated Love number and tidal heating rate, the rates of change of $e$, $a$, $T_m$ and $T_c$ can be determined, following which the internal structure and thermal properties are updated. Continuing this process, evolutional calculations based on the coupled calculations become possible. In our calculation, we integrate each value forward with a time step of $10^5$ years for 10 Gyr. We tried other time step and confirmed that the original time step is valid for calculations. In some cases, mantle temperature exceeds the core temperature because the mantle is heated by tidal heating. In this case, we calculate the heat flux from the mantle to the core following Nimmo \& Stevenson (2000). We separate the core into spherical shell and solved the conduction equation with 10$^4$ years of time step. However, if the stagnant lid thickness becomes very large, the convection in the mantle stops. Compared with the convective layer, the temperature of the conductive layer is very low. Thus, the magnitude of the tidal heating is much smaller than that of a convective planet, and the tidal heating has little effect on the evolution. Our simple thermal and structural models are not valid for a conductive planet. In this work, the tidal heating in a conductive structure is out of our scope. Thus, when the convective mantle thickness becomes zero, we stop the calculations. Integrations were performed using the Runge-Kutta method. As for the initial values of eccentricity $e_0$, we considered a range of values less than 0.5. The initial values of $T_m$ and $T_c$ are not constrained. Thus we tried a wide range of temperatures between 1600 K and 2000 K. In addition, because the tidal heating in the mantle strongly depends on the mantle viscosity, we consider the reference viscosity $\eta_0$ as a free parameter between $10^{19}$ Pa s and $10^{20}$ Pa s.

\section{Results}
\begin{figure}[htbp]
\centering
\includegraphics[width=13cm] {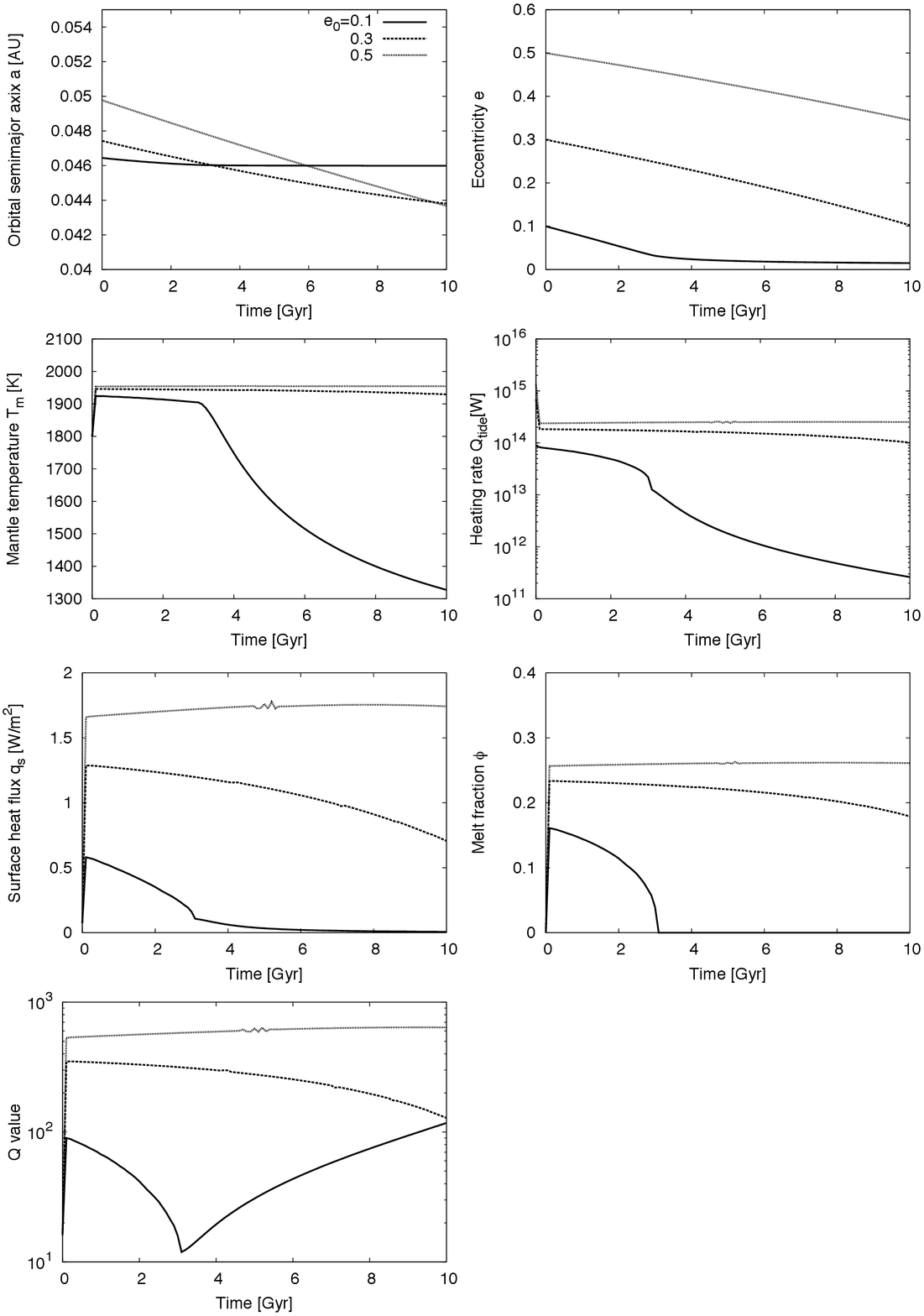}
\caption{Orbital and thermal evolutions of the planet. Mass of the central star $M_{\ast}$ is 0.1$M_{\odot}$. $\eta_0=10^{19}$ Pa s. Initial values of the mantle temperature and core are 1800 K. Initial value of the eccentricity is 0.1 (solid line), 0.3 (dashed line) and 0.5 (dotted line).}
\label{fig3}
\end{figure}

\begin{figure}[htbp]
\centering
\includegraphics[width=13cm] {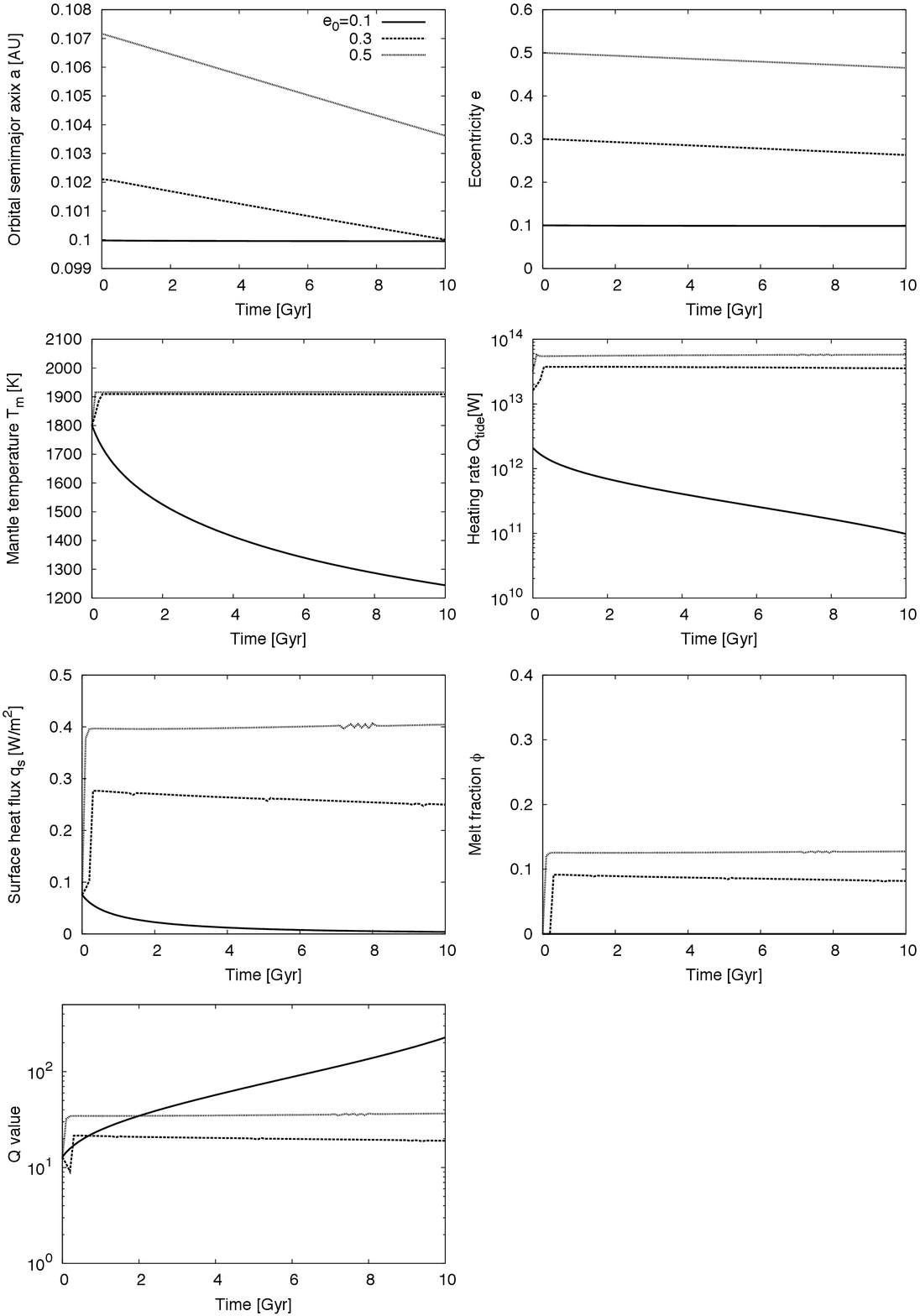}
\caption{Orbital and thermal evolutions of the planet. Mass of the central star $M_{\ast}$ is 0.2$M_{\odot}$. $\eta_0=10^{19}$ Pa s. Initial values of the mantle temperature and core are 1800 K. Initial value of the eccentricity is 0.1 (solid line), 0.3 (dashed line) and 0.5 (dotted line).}
\label{fig4}
\end{figure}

The thermal and orbital evolutions of the three-layer model are shown in Figs. \ref{fig3} and \ref{fig4}. The initial values of $T_m$ and $T_c$ were 1800 K. The initial value of the semi-major axis was calculated considering habitability with 0\% cloud cover. Surface temperature is determined by Eq. (\ref{Ts}). In the case of $M_{\ast}=0.1M_{\odot}$ (Fig. \ref{fig3}), two types of evolution can be seen, depending on the eccentricity. Due to the initial conditions, a heating rate of more than $10^{14}$ W was generated with every eccentricity. Because of this large heat, the mantle temperature increases, inducing runaway heating. When the temperature exceeds the solidus temperature, the mantle begins to melt and the heating rates decrease as shown in Fig. \ref{heat}. Because of the melt, runaway heating stops, and an equilibrium state is reached in which the radiated-heat rate is equal to the tidal heating. If $e_0$ is large enough, large heat is generated as shown in Eq. (\ref{tide}). Thus, the melt fraction becomes larger, which results in a large value of Q. Due to the large Q value, the eccentricity and semi-major axis do not change significantly even though $M_{\ast}$ is small (the habitable zone is close to the star). 
Although semi major axis decreases by the tidal dissipation, it is far from the inner edge of the habitable zone in every case, which means that the planet stays in the habitable zone for more than 10 Gyr. 

One caveat is that habitable zone itself changes with time because the luminosity of the central star changes. Thus, our estimation for the stability at the habitable zone of the planet may be too rough. However, Kopparapu et al. (2013) calculate the "continuous" habitable zone in which planets are habitable for 5 Gyr. In their calculation, continuous habitable zone of 0.1$M_{\odot}$ mass star is between 0.2 AU and 0.5 AU. They also show the continuous habitable zone based on the model by Selsis et al. (2007). Our calculation results of the semi major axis are in the range of the continuous habitable zone of both models. Thus, we can say that Martian-sized planets can stay in the habitable zone for at least  5 Gyr. 

If the value of Q is constant of the order of 10, the eccentricity is rapidly reduced and the tidal heating does not largely affect the thermal evolution of the planet. The orbital distance also changes significantly. However, considering the melt, the Q value increases and the large eccentricity and semi-major axis are maintained, which is a difference between constant Q value calculations and the coupled calculations. 

Due to the large eccentricity, relatively large heat generation occurs and the high temperature is maintained. This tidal heating is not too large because the melt reduces the viscosity and shear modulus. Thus, an equilibrium state occurs. 
However, at $e_0=0.1$, the heating rate is relatively small, which reduces the tidal heating, making an equilibrium state impossible to maintain. When the heating rate is reduced, runaway cooling occurs and the melt fraction also decreases. Radiogenic heating is not sufficient to heat up the mantle. The decreasing melt fraction causes the Q value to drop drastically. However, once the melt fraction becomes zero, the Q value increases again because the mantle temperature drops. The results in Fig. \ref{fig3} are for the case in which the initial value of $T_c$ and $T_m$ were both 1800 K. We also tried with $T_c=3000$ i.e., with an initial temperature difference of 1200 K. In that case the core temperature rapidly reaches the mantle temperature. Thus, the effect of a large $T_c$ is small in terms of the evolution of the planet. 

Barnes et al. (2009) analyzed the habitability focusing on the surface heat flux. Conventionally, the habitable zone is defined as the area where H$_2$O is kept in its liquid phase from models which assume that the ground is water saturated (e.g., Kasting et al. 1993). However, as well as liquid water, the surface state also plays an important role in determining habitability. Barnes et al. (2009) gave a condition for habitability based on the heat fluxes and surface states of Mars and the Jovian satellite Io. In their estimation, a heat flux of between 0.04 W m$^{-2}$ and 2 W m$^{-2}$ is required for habitability. If the magnitude of the heat flux is similar to Io, active volcanoes make the surface inhabitable. In our calculations, the hot equilibrium state of the planet radiates more than 0.5 W m$^{-2}$ when $M_{\ast}=0.1M_{\odot}$ and $e_0$ is 0.3, which is the suitable value for the habitability. At $e_0$=0.5, although it does not exceed 2 W m$^{-2}$, the surface heat flux reaches the maximum limit. Thus, the planet is predicted to have active volcanos like Io, which may make the planet inhabitable. If $e_0$ is less than 0.1, tidal heating is not enough to maintain the melt in the mantle, which results in cooling of the planet. From the coupled calculations, we can say that a Martian-sized planet that orbits in the habitable zone of an $M_{\ast}=0.1M_{\odot}$ star is suitable for the habitability if the initial eccentricity is between 0.2 and 0.5. However, when the eccentricity is around 0.1, the heat flux exceeds 0.5 W m$^{-2}$ at $e_0$=0.1. Thus, when we observe the Martian sized planet at the habitable zone of 0.1M$_{\odot}$ star, it is important whether the planet has the eccentricity more than 0.1 for the habitability. If observed eccentricity is over 0.1, the planet is predicted to have partial melt and the relatively large heat flux.

Figure \ref{fig4} shows the case for $M_{\ast}=0.2M_{\odot}$. Compared to the star with $M_{\ast}=0.1M_{\odot}$, the melt fraction is small. At $e_0=0.1$, melt is not generated. Because the planet is assumed to orbit according to Kepler's law, the mean motion in the habitable zone is small. As shown in Eq. (\ref{tide}), the heating rate increases with $n^5$. Thus, tides cannot produce enough heat to generate large amount of melt in the case of $M_{\ast}=0.2M_{\odot}$ star.
Surface heat flux at the hot state is around 0.3 W m$^{-2}$ and 0.4 W m$^{-2}$ when $e_0$ is 0.3 and 0.5, respectively, which is suitable range for the habitability.
The eccentricity does not change significantly for $M_{\ast}=0.2M_{\odot}$ even though the Q value is relatively small because the orbital distance is large, which is different from the case of a $0.1M_{\odot}$ mass star. In the case of $M_{\ast}=0.1M_{\odot}$, the orbital distance is small. However, the small rate of change of eccentricity is caused by the large $Q_p$ due to the melt. 

Through the coupled calculations, it has been determined that the thermal state of small extrasolar planets in habitable zones depends mainly on their initial eccentricity and the mass of their star. The initial value of the core temperature does not affect the heating state because tidal heating soon heats up the mantle anyway. Figure \ref{phase} shows the thermal state as a function of the mass ratio of the star ($M_{\ast}/M_{\odot}$) and the initial eccentricity $e_0$ for different initial temperatures ($T_{m0}$) and reference viscosities ($\eta_0$). As can be seen in Fig. (\ref{phase}), the reference viscosity and initial temperature do not affect the heating states of the planet. If the mass of the star is around 0.1$M_{\odot}$, the planet reaches a hot state when $e_0>0.2$,  independently of $T_{m0}$ and $\eta_0$. As mentioned above, the planet can stay in the habitable zone for a long time in the hot state. If the mass of the star is around 0.2M$_{\odot}$, a hot state occurs if $e_0>0.3$. If $M_{\ast}/M_{\odot}$ is more than 0.3, a hot state is not formed even though the eccentricity is quite large. The habitable zone of a star in this mass range is too far from the star for tidal heating large enough for runaway heating and melting of the mantle to occur. Since the cold planet has low temperature and rigid, stars of less than $0.2M_{\odot}$ in mass are good targets for finding habitable planets. 

\section{Discussion}
\subsection{Morphology}
If the tidal heating is sufficiently large, the mantle is heated and the surface heat flux exceeds the Martian heat flux. In this work, we assume stagnant lid convection as a heat transport mechanism. As an alternative  mechanism, plate tectonics may occur in planets in a hot state. Although it is currently only possible to observe the plate tectonics of Earth, it has been suggested that plate tectonics occurs in large, rocky extrasolar planets (e.g., Valencia et al. 2007). However, plate tectonics is a very complicated system and the main induction mechanism is not well understood. The size of the planet may affect the stress (Valencia et al. 2007; O'Neill \& Lenardic 2007); however, Korenaga (2010) suggests that the effect of surface water is more important than the size of the planet. In our model, we do not include surface oceans and the size of the planets considered is relatively small.  Thus, stagnant-lid convection is more reasonable. However, in the case of Mars, plate tectonics has been suggested to be the origin of the early internal dynamo (Nimmo \& Stevenson 2000). Thus, plate tectonics cannot be ruled out for Martian-sized extrasolar planets. 

Delamination may be more probable in small-sized extrasolar planets than plate tectonics. In a hot state, the melt in the mantle reduces the mantle viscosity. Thus, the bottom of the stagnant lid is delaminated and absorbed into the mantle. Delamination displaces the surface of the planet, which is the suggested cause of the corona of Venus (Smrekar \& Stofan 1997).  Thus, we can suggest that the planet is heated by tides, and the morphology of the planet is mainly a product of localized delamination processes caused by partial melting. As an alternative heat transfer model, heat pipe mode is suggested in Io and the Earth (Moore 2001; Moore \& Webb 2013), which can be the important heat transfer mechanism for the tidally heated extrasolar planets. Thus, the effects of delamination and heat pipe should be considered in future work.

\subsection{Magnetic field}
It was revealed that Mars had an intrinsic magnetic field in the past (Acu\~{n}a et al. 1999). The most probable origin of the intrinsic magnetic field is the dynamo in the core. The mechanism of the Martian dynamo is not well understood. However, it is required that more heat than that transported by adiabatic heat flux is transported from the core to the mantle in order to induce the core convection (Nimmo \& Stevenson 2000).  If tidal heat is not generated in the mantle, heat is transported sufficiently by the mantle and a dynamo may be induced. However, in our calculations, tidal heating heats up the mantle. In some cases, heat is even transported from the mantle to the core. Thus, in the hot state of a planet, it is unlikely the dynamo is activated. 

An intrinsic magnetic field is not predicted in a tidally heated planet; however, if the orbital plane of the planet is oblique to the magnetic axis of the star, an induced magnetic field may be possible because of the melt in the mantle. In the case of Io, an induced magnetic field is observed (Khurana et al. 2011) . The induced magnetic field of Io is consistent with a mantle of at least 20 \% partial melt (Khurana et al. 2011). This melt fraction is comparable to our calculation results.  

\subsection{Heterogeneity of the planet}
For simplicity, we assume that the internal structure of the planet is spherical shells and depends only on the radial distance. As for the habitable zone, the spherical-shell structure should provide a valid approximation because the temperature in light side is not so high. However, the heterogeneity of the planet is an interesting topic for future study, especially for planets that orbit significantly close to their stars. Tidal heating in a planet with a heterogeneous structure was calculated for the case of Enceladus, though it is extremely complicated (Tobie et al. 2008; B\v{e}hounkov\'a et al. 2012). If coupled calculations including a heterogeneous structure are possible in future work, an important comparison to the spherical-shell model of this work can be made. 

\subsection{Resonance}
One caveat is that the coupled calculations in this work do not consider the resonance of the planet. If the planet is in a resonant state with other planets, the orbital evolution becomes significantly more complicated. Resonance among planets increases the planets' eccentricity, as exhibited by the Jovian satellites Io, Europa and Ganymede (Murray \& Dermott 1999). Some satellites around Saturn also interact with each other (Murray \& Dermott 1999). Hence, if there are multiple planets around a star, we must consider the resonance between them. 

\section{Summary}
\begin{figure}[htbp]
\centering
\includegraphics[width=13cm] {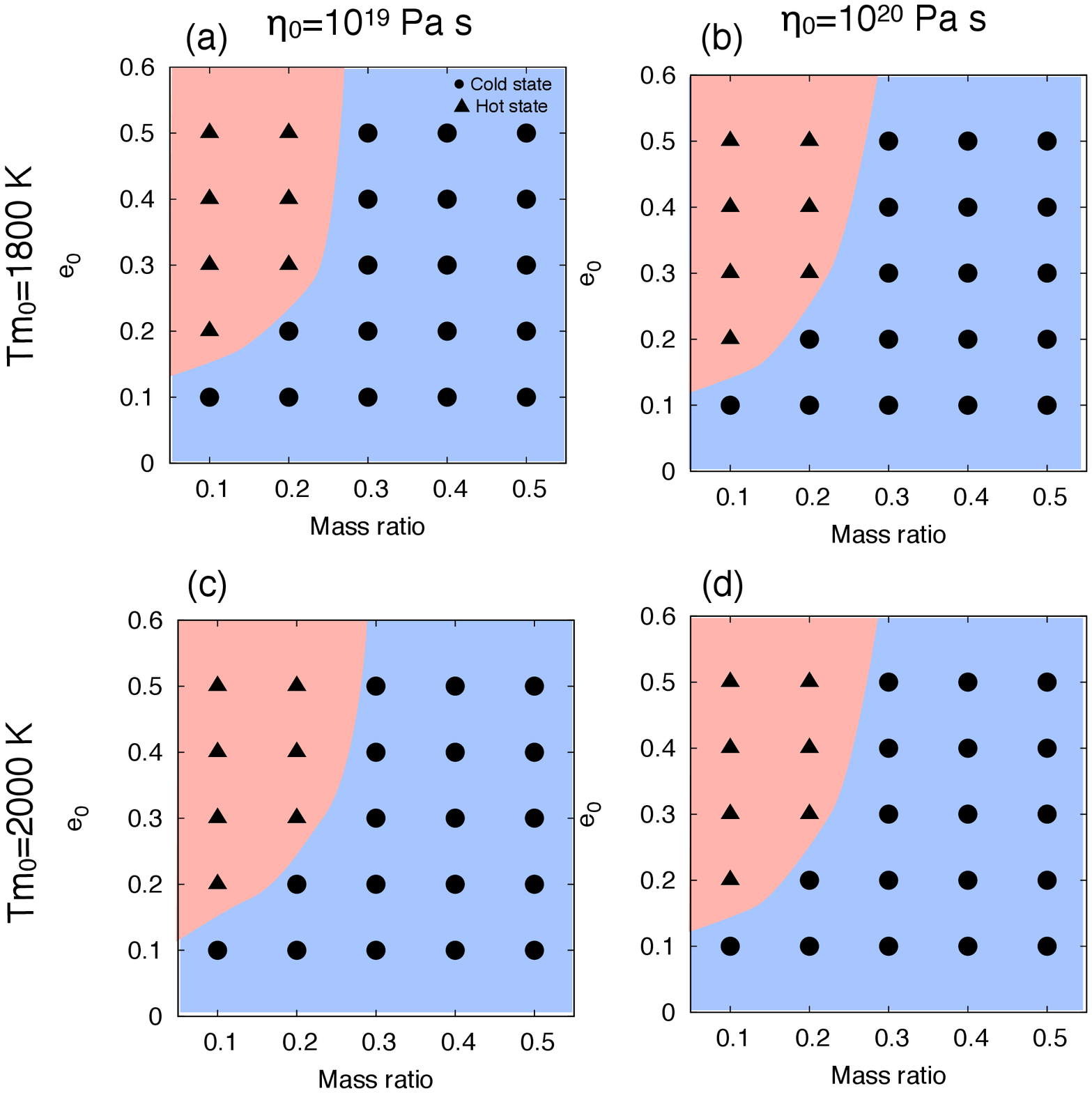}
\caption{Heating state of the planet as a function of mass ratio of the star ($M_{\ast}/M_{\odot}$) and initial eccentricity of the planet ($e_0$). Triangle points show the hot state in which runaway heating and the melt induced equilibrium state occur. Circle points show the cold state, in which runaway cooling occurs. Top panels (a) and (b) show the results for $T_{m0}$=1800 K. Bottom panels (c) and (d) show the results for $T_{m0}=2000$ K. $T_{c0}$ is the same as $T_{m0}$. In the left panels (a) and (c), $\eta_0=10^{19}$ Pa s, and in the right panels (b) and (d) $\eta_0=10^{20}$ Pa s. }
\label{phase}
\end{figure}

Through thermal and orbital coupled calculations, it was found that two types of evolution are probable for Martian-sized planets in the habitable zones of M stars. One type is the runaway cooling state in which the radiated heat flux is larger than the tidal heating, decreasing the mantle temperature through positive feedback. As a result of the cooling, convection in the mantle is reduced and the planet becomes rigid. The other type of evolution is the hot state. In a hot state, tidal heating exceeds the radiated heat flux and runaway heating occurs. However, melt in the mantle reduces the tidal heating by decreasing the viscosity and shear modulus, and an equilibrium state is reached. The eccentricity and orbital distance of hot state planets do not change significantly due to the large Q value caused by the melt. Thus, the planet in hot state can stay the habitable zone for more than a few billion years. In the equilibrium state, the heat flux is in the range suitable for habitability (Barnes et al. 2009). We performed the calculations and by changing many parameters, found that those that have the most significant effect on the evolution are the mass of the star and the initial value of the eccentricity. If the mass of the star is less than 0.2$M_{\odot}$ and the initial eccentricity is more than 0.2, the planet should be in a hot state (Fig. \ref{phase}). A hot state planet contains melt in the mantle; thus, a magnetic field may be induced. An intrinsic magnetic field, however, is unlikely to occur because the mantle temperature is large and sufficient heat is not transported from the core.

Among extrasolar planets, super-earths in particular are considered for habitability. In the case of a super-earth, our simple structural model is not valid and the effects of pressure are likely large (Wagner et al. 2012).  We must also consider the phase of the core because the core may be solidified by the large pressure. However, our simple coupled calculations provide the first step towards these more complicated and accurate analyses of extrasolar planets.

\acknowledgments
This work is supported by JSPS Research Fellowship. V. Dobos gave us useful comments about the viscosity of the planet.

\clearpage



\clearpage

\begin{table}[htbp]
\centering
\caption{Physical parameters and values.}
\label{value}
\begin{tabular}{lcccc} \hline
Parameter&Symbol&Value&Unit&Reference\\ \hline
Planet radius&$R_p$&3390&km&1\\
Planet mass&$M_p$&$6.3\times10^{23}$&kg&1\\
Core radius&$r_c$&1550&km&1\\
Lid density&$\rho_l$&3500&kg m$^{-3}$&\\
Mantle density&$\rho_m$&3500&kg m$^{-3}$&1\\
Core density&$\rho_c$&7200&kg m$^{-3}$&1\\
Mantle heat capacity&$c_m$&1200&J kg$^{-1}$ K$^{-1}$&2\\
Core heat capacity&$c_c$&840&J kg$^{-1}$ K$^{-1}$&2\\
Latent heat of melting&$L_m$&$6\times10^5$&J kg$^{-1}$&3\\
Ratio of mean and upper mantle temperature&$\epsilon_m$&1.0-1.05&&1, 3\\
Ratio of mean and core temperature&$\epsilon_c$&1.1&&1, 3\\
Lithosphere thermal conductivity&$k_l$&4.0&W m$^{-1}$ K$^{-1}$&\\
Mantle thermal conductivity&$k_m$&4.0&W m$^{-1}$ K$^{-1}$&1\\
Core thermal conductivity&$k_c$&50.0&W m$^{-1}$ K$^{-1}$&2\\
Mantle expansivity&$\alpha$&$4\times10^{-5}$&K$^{-1}$&2\\
Core expansivity&$\alpha_c$&$2\times10^{-5}$&K$^{-1}$&2\\
Mantle diffusivity&$\kappa$&10$^{-6}$&m$^2$ s$^{-1}$&1\\
Albedo&$A$&0.3&&4\\
Reference temperature&$T_0$&1600&K&5\\
Mantle activation energy&$E$&300&kJ mol$^{-1}$&2\\ 
Lid shear modulus&$\mu_l$&50&GPa&5\\
Lid viscosity&$\eta_l$&10$^{40}$&Pa s\\
Anelasticity parameter&$n_a$&0.25&&6\\
Melt fraction coefficient&$B$&30&&7\\  \hline
\end{tabular}
\tablerefs{
(1) Grott \& Breuer 2008a; (2) Nimmo \& Stevenson 2000; (3) Breuer \& Spohn 2006; (4) Paige et al. 1994; (5) Fischer \& Spohn 1990; (6) Jackson et al. 2002; (7) Moore 2003
}

\end{table}

\end{document}